\documentclass[preprint]{aastex}

\def\ltsim{ \,{}^<_\sim\, }
\def\gtsim{ \,{}^>_\sim\, }

\shorttitle{NGC 5128 Inner Halo}
\shortauthors{Harris \& Harris}

\begin{document}

\title{The Halo Stars in NGC 5128. III:  \\
An Inner-Halo Field and the Metallicity Distribution\altaffilmark{1}}

\author{William E.~Harris}
\affil{Department of Physics \& Astronomy, McMaster University,
    Hamilton ON L8S 4M1\altaffilmark{2} }
\email{harris@physics.mcmaster.ca}

\and

\author{Gretchen L.~H.~Harris}
\affil{Department of Physics, University of Waterloo, Waterloo ON N2L 3G1\altaffilmark{2}}
\email{glharris@astro.uwaterloo.ca}

\altaffiltext{1}{Based on
Observations with the NASA/ESA {\sl Hubble Space Telescope}, obtained at
the Space Telescope Science Institute, which is operated by the Association
of Universities for Research in Astronomy (AURA), Inc., Under
NASA Contract NAS 5-26555.}

\altaffiltext{2}{Visiting Fellow, Research School of Astronomy and 
Astrophysics, Australian National University, Weston ACT 2611, Australia}

\begin{abstract}

We present new {\it Hubble Space Telescope}
WFPC2 $(V,I)$ photometry for field stars in NGC 5128
at a projected distance of 8 kpc from the galaxy center, 
which probe a mixture of its inner halo and outer bulge.
The color-magnitude diagram shows an old
red-giant branch which is even broader in color than our two
previously studied outer-halo fields (at 21 and 31 kpc), with significant numbers
of stars extending to Solar metallicity and higher.  The peak frequency
of the metallicity distribution function (MDF) is at [m/H] $\simeq -0.4$, with
even fewer metal-poor stars than in the outer-halo fields.  If we use
the 21- and 31-kpc fields to define template ``halo'' MDFs and subtract
these from the 8-kpc field, the residual ``bulge'' population has a
mean [m/H] $\simeq -0.2$, similar to the bulges of other large spirals
and ellipticals.
We find that the main features of the halo
MDF can be reproduced by a simple chemical evolution model in which
early star formation goes on simultaneously with an initial stage of
rapid infall of very metal-poor gas, after which the infall 
dies away exponentially.
Finally, by comparison with the MDFs for the NGC 5128 globular clusters,
we find that in all the halo fields we have studied, there is
a clear decrease of specific frequency $S_N$ (number of clusters per unit halo
light) with increasing metallicity.  At the lowest-metallicity range
([Fe/H] $< -1.6$) $S_N$ is $\sim 4 - 8$, while 
at metallicities [Fe/H] $> -1$ it has dropped to $\simeq 1.5$.  
This trend may indicate that
globular cluster formation efficiency is a strong function of the 
metallicity of the protocluster gas.
However, we suggest an alternate possibility, which is that globular
clusters form preferentially sooner than field stars.
If most of the cluster formation within a host giant molecular cloud (GMC)
takes place sooner than most of the distributed field-star formation, and
if the earliest, most metal-poor star-forming clouds are prematurely disrupted 
by their own first bursts of star formation, then
they would leave relatively few field stars with a high$-S_N$ population.
The high specific frequency at low metallicity may therefore be related to the
comparably large $S_N$ values found in the most metal-poor dwarf ellipticals.  

\end{abstract}


\section{Introduction}

An important part of the historical record of galaxy formation is
locked up in the ages and metallicities of the earliest stellar populations,
particularly those found in the halos of galaxies.
For most galaxies, however, even the brightest old-halo stars are too dim
to be seen as anything but a smooth distribution of integrated light, from
which it is difficult to deduce much more than the
mean age and metallicity of the population \citep[e.g.,][]{tra00}.
Giant ellipticals present a special challenge because they are the only
galaxy type not found in the Local Group, thus few are 
close enough to allow direct, comprehensive probes 
into their stellar populations.

NGC 5128, the central giant elliptical in the Centaurus group
just 4 Mpc distant, is an outstanding exception.
Recently, we used deep $(V,I)$ photometry from $HST/WFPC2$ to obtain
color-magnitude diagrams for fields  in the outer halo of NGC 5128, at
projected galactocentric distances of 21 and 31 kpc
\citep[][hereafter Papers I and II]{har99,har00}.
The CMDs of these halo fields reveal the top $\sim2.5$ magnitudes of a 
normal old red-giant branch (RGB), with a wide 
range in color that is indicative of a broad metallicity range.  Since 
the RGB loci for old stars are predominantly sensitive to metallicity rather
than age, they can be used to construct
a first-order {\it metallicity distribution function} (MDF) 
-- the first such one for any
giant elliptical and one which does not rely on indirect arguments from
luminosity-weighted integrated light.  The MDF is an important step
towards constructing an eventual age-metallicity relation (AMR)
to constrain its evolutionary history.

In the present study, we present similar data for an ``inner halo''
(or alternatively an ``outer bulge'') field much closer in to the
center of NGC 5128, at a projected distance of 8 kpc.  This field reveals
important differences in the MDF compared with the outer-halo fields
and, perhaps, give us the first
direct look at the pure bulge population of a giant elliptical.

\section{Observations and Data Analysis}

The location of our 8-kpc target field is southwest of the center of NGC 5128
with the PC1 chip center at
$13^h 24^m 53^s$, $-43^o 04' 35''$ (J2000).
This location avoids the central dust lane (see Figure 1 of Paper II 
for a finder chart), and comparison with a deep $(B-R)$ color image by
\citet{pen01} sensitive to the presence of dust shows that it also avoids
two thin dust streamers extending southward along the major axis.
The raw observations, taken in HST Cycle 8 for program GO-8195, consisted of
a sequence of 14 half-orbit exposures in F606W (wide $V$) totalling 17500 seconds,
and 10 exposures in F814W (wide $I$) totalling 12100 seconds, the same
as for our 31-kpc field taken during the same program.

Data analysis paralleled what we did for the 31-kpc field and our procedures
are discussed in detail in Paper II.
Pointings in the sequence of raw images differed by several pixels in 
($\Delta x, \Delta y$), and we reregistered the individual exposures
and median-combined them with the normal IRAF and STSDAS
packages to produce clean median images free of
cosmic rays, bad pixels, and other artifacts.
Photometry on the master $V$ and $I$ frames was then carried out with
the DAOPHOT II code \citep[][]{ste92} in the same manner as described
in Papers I and II, with a two-pass sequence of FIND, PHOT, and ALLSTAR.  
Transformation of the instrumental magnitudes 
to final $V,I$ values followed the prescriptions of \citet{hol95}.

The one important respect in which the data reduction for this field
differed from our previous work arose from the level of stellar crowding.
While the two outer fields (Papers I, II)
are completely uncrowded at the characteristic
WFPC2 $0\farcs1$ resolution, the 8-kpc field
has a sufficiently high density of resolved stars to make the photometry
in the three WF chips quite difficult and any discussion of
these is not pursued further here.  On the PC1 chip
with its superior resolution, the photometry is much more easily
manageable with careful attention to defining the stellar point spread
function (PSF).  We carried out complete reductions of the PC1 images
with five different empirically constructed PSFs:
(1) the Gaussian-plus-residuals model in DAOPHOT, constructed from an average of
several bright stars in the PC1 chip with three iterations of ``cleaning'' 
neighbors from around the PSF candidate stars, (2) the Lorentzian-plus-residuals
model in DAOPHOT,
(3) PSFs for $V$ and $I$ supplied from the STScI archive, (4) PSFs
derived from uncrowded bright stars on other PC1 images (Stetson, private
communication), and (5) PSFs taken from the PC1 images in
our uncrowded outer-halo fields.
In each case we independently rederived the necessary aperture corrections
through curves of growth
to normalize the PSF-fitted magnitudes to the $0\farcs5$ aperture radius
used in the Holtzman et al.~calibrations.

The results which clearly gave the cleanest fits to the image, and which we
adopted, were the 
Lorentzian-based PSFs constructed iteratively from the 8-kpc PC1 field 
itself.  However, despite differences in the PSF profile shapes and 
the consequent aperture corrections, the final lists of calibrated $(V,I)$
stellar magnitudes (essentially resting on 
the PSF scalings derived by ALLSTAR) proved to be very similar across
all five types of PSF models.

The final color-magnitude diagram (CMD) in $(I,V-I)$ for a total of
17,283 stars on PC1 detected in both filters is shown in Figure \ref{fig1}.
The full dataset can be obtained from the authors on request.
For this inner field, we made no attempt to find or remove nonstellar
objects from the list; although there are a few faint background galaxies
visible in the field, their numbers (from the image classification
analysis done more completely in Papers I and II) are 
quite small here compared with the overwhelmingly large population
of stars, and the high degree of crowding would, in any case, vitiate objective
measurements of nonstellar image shapes for most such objects.

The CMD, when compared with our previous data for the outer-halo fields,
shows the same old RGB population with a bright-end
``tip'' at $I\simeq 24.2$,
but extends even further to the red than did the outer fields.
The number of stars appearing above the RGB tip ($I < 24$) that might arise
from a younger AGB population, for example \citep[see][and Papers I and II]{sor96},
is statistically negligible.  Within the main RGB distribution, however,
our deductions about the range of metallicities of these stars
depend on the effects of photometric measurement scatter,
and the much higher level of crowding here makes this question more acute.
To investigate the photometric scatter, we carried out several simulations
in which 500 stars at a time were added to the original images, with input 
colors and magnitudes following an ideal, narrow RGB sequence.
These images were then measured in exactly the same procedure
(two-pass FIND/ALLSTAR), yielding the results shown in Figure \ref{fig2}.

The artificial-star tests show higher scatter than
did the same tests on the outer fields (as expected from the higher degree of
crowding; compare Figure 5 from Paper II), but within the upper part 
of the RGB, the observed color spread is far larger than can be explained
by photometric errors and must be intrinsic to the stellar population.
The trend of measurement scatter in each band is shown in Figure \ref{fig3}.
Encouragingly, in both $V$ and $I$ the simulations revealed 
no photometric {\it bias}
larger than $\sim 0.03$ mag at any level of interest here, 
i.e. much smaller than the level of measurement uncertainty.  
We therefore made no attempts to apply any bias corrections. 

Lastly, the artificial-star tests were used to estimate the photometric
detection completeness as a function of magnitude.  The levels at which
the numbers of detected stars drop to 50\% of the numbers put in can be
expressed (once the instrumental $(v,i)$ magnitudes are folded through
the Holtzman et al.~calibration relations) as
$$I_{50} = 26.84 - 0.062 (V-I) + 0.025 (V-I)^2 \, \, {\rm for} \, I ,$$
$$I_{50} = 28.76 -1.247 (V-I) + 0.065 (V-I)^2 \, \, {\rm for} \, V . $$

Approximate completeness curves in $V$ and $I$ are shown in Figure \ref{fig4}. The
two smooth curves are Pritchet interpolation functions \citep{fle95}
of the form 
$$ f = 0.5 \Big( 1 + {{a (m-m_0)} \over {\sqrt{1 + a^2 (m-m_0)^2}}}\Big) \,  $$
where $m_0$ is the magnitude at which 50\% completeness is reached
and $a$ is a parameter governing the steepness of the decline in $f$.
The curves shown in the Figure are for a 
particular color $(V-I) = 2$ and are only to be taken
as indicative of the average magnitudes at which the photometry becomes
severely incomplete. In practice, we define $f$
in terms of the {\sl instrumental} magnitudes $(v,i)$ and then
translate them through the photometric calibration equations
into the appropriate $f(V,I)$ for each point in the CMD.

For the purposes of deriving the
stellar metallicity distribution, detection incompleteness is
unimportant {\sl except} at the extreme upper right corner of the CMD, 
where the very reddest
and brightest RGB stars fall near the completeness cutoff determined by
the $V$ filter.  As will be seen below, these reddest
stars are likely to have above-Solar heavy-element abundances; they are
redder than the old-RGB populations in any previously observed globular clusters,
dwarf galaxies, or halos of larger galaxies.
Despite our deliberately planned exposure times in $V$, which were considerably
longer than in $I$ (and our usage of the
``wide $V$'' F606W filter rather than the more nearly standard F555W), we 
may still be missing some fraction of the very reddest RGB stars.  
In the discussion below on the metallicity distribution, we explicitly
include the effects of incompleteness as far as the data allow.
However, it should be kept in mind that if the outer bulge of
NGC 5128 does contain stars well above
Solar metallicity, our data are not capable of detecting most of
them.  Such objects,
if they exist, will have to be searched for with longer-wavelength $JHK$
imaging, such as with the NICMOS camera \citep[see][]{mar00}.

\section{The Metallicity Distribution Function}

\subsection{Calibrating the Metallicity Measurement}

The derivation of stellar metallicities follows the procedure
developed in Paper II:  briefly, we superimpose a stellar model grid
on the CMD and interpolate within it to calculate a heavy-element
abundance $Z$ for each star.  Our adopted metallicity index is then
by definition [m/H] = log $(Z/Z_{\odot})$.  The main stellar model grid is
that of \citet{van00}, which provides fiducial RGB evolutionary tracks
in metallicity intervals of roughly 0.1 dex in log $Z$ from $Z = 0.00017$ up to
$Z = 0.013$.  Solar metallicity is adopted as $Z_{\odot} = 0.017$.
However, a significant fraction of the stars in our inner-halo
sample are more metal-rich than this, and particular attention needs to be 
paid to extending the grid to higher $Z$.  Unfortunately, it
is precisely this high$-Z$ regime which is fraught with the greatest uncertainty
in every direction.  In Paper II, we used two metal-rich isochrone
lines from \citet{ber94} ($Z = 0.02, 0.05$) to extend the grid.  However, these have
scaled-Solar element ratios and are thus fundamentally not compatible with
contemporary $\alpha-$enhanced models.  This issue had little relevance
for the outer-halo fields (Papers I and II) because these 
fields did not have many stars
which extended into the high$-Z$ range, but for our inner-halo data the
question is more pressing.  Stars with [$\alpha$/Fe] $> 0$ are hotter than
ones with scaled-Solar ratios because of their lower opacities, and simple
renormalization of the total $Z$ becomes risky particularly at high
metallicity \citep[][]{van00,sal98}.

Ideally, we would like to employ a grid of RGB models which is 
(a) computed with full $\alpha-$enhancement,
(b) extends to at least $3 Z_{\odot} \simeq 0.05$,
(c) is densely spaced in log $Z$,  and (d) 
also properly accounts for any progressive change in [$\alpha$/Fe] with
metallicity itself \citep[e.g.][]{she01}.  Such material is not yet available.
Recently \citet{sal00} have published metal-rich and $\alpha-$enhanced
isochrones for a widely spaced grid of
heavy-element values $Z = (0.008, 0.019, 0.04, 0.07)$. However, at least
in the upper-RGB luminosity range which we require, the theoretical $(V-I)$ colors
of all these tracks are significantly bluer for a given metallicity
compared with the \cite{van00} models, as well as (more importantly) 
the real star clusters that we use
to calibrate the grid (see below).  A more finely spaced grid of metal-rich
models is available from \citet{yi01}, but these are calculated 
for scaled-Solar abundances.
For the present we are forced to adopt an alternate procedure.

Our approach is a continuation of the one we adopted in Paper II:  instead of relying solely
on theoretically computed color indices, we 
pin the model tracks as closely as possible to the
RGB loci and abundances of real Milky Way clusters.  At low $Z$, this method
works handily because there are many well observed globular clusters with
low metallicities, low reddenings, and accurate RGB loci in the $(M_I, V-I)$ plane.
The five which we use to cover the range from [m/H] $= -2.0$ up to
[m/H] $= -0.4$ are M15, NGC 6397, NGC 6752, NGC 1851, and NGC 104 (47 Tuc).
At higher $Z$, the situation is considerably worse.  The two metal-rich objects we rely
on, out of sheer necessity, are the bulge globular cluster NGC 6553 and the 
very old disk cluster NGC 6791.  NGC 6553 is generally thought to have
essentially Solar heavy-element abundance \citep{bar99,coh99,car01,bea01,ori01}
with [Fe/H] $\simeq -0.2 \pm 0.1$ and [$\alpha$/Fe] $\simeq 0.2 - 0.3$.
NGC 6791 is generally thought to have [Fe/H] $\simeq +0.4$ with little or no
$\alpha-$enhancement \citep{pet98,cha99}, but see \citet{tay01}
for a healthy discussion of concerns.  Both clusters have observational
handicaps that are much more serious than in the low$-Z$ regime:  NGC 6553
is both heavily reddened and differentially reddened, while NGC 6791 has
an age which is younger than the standard globular clusters by $\sim 2$ Gy
\citep[e.g.][]{cha99}.  To work around these difficulties, we have taken 
the Yale scaled-Solar tracks \citep{yi01} 
and have used them to define the {\sl shapes} of
the tracks in the $(M_I, V-I)$ plane 
near $Z=Z_{\odot}$.\footnote{The VandenBerg et al. (2000) models use
$Z_{\odot} = 0.017$ while the Yale models use $Z_{\odot} = 0.018$.
At the level of accuracy needed for our purposes, this difference is negligible.}
For NGC 6791, we adopt a fiducial value $Z = 0.045$ and use the models to
find that we must shift its upper RGB 
0.15 mag redder in $(V-I)_0$ to correct it to an age 2 Gy older.
This adjusted locus, by definition, is the uppermost boundary of the metallicity
sequence we can use at present.  Lastly, we also interpolate within the
models to define a track at $Z = 0.025$, in between NGC 6553 and NGC 6791.

The complete array of fiducial tracks and star clusters that we have adopted 
is shown in Figure \ref{calibration}.  The fiducial points for NGC 6553, 
for which $Z \simeq 0.017 = Z_{\odot}$, fall slightly redder 
than the most metal-rich VandenBerg et al.\ track at $Z = 0.013$, and
the raw NGC 6791 sequence (not corrected for age differences) lies as it
should between the tracks for $Z = 0.025$ and 0.045.  It is obvious that
these observational constraints on the high$-Z$ part of
the grid, from NGC 104 up to NGC 6791, are very thin indeed and that our
current definition of this part of the grid is only a stopgap measure.  With further
careful observations of selected globular clusters, it may be possible
to add one or two more fiducial sequences in this range, 
but particularly urgent attention is needed to the computation of
appropriate models.

We emphasize once again that {\sl the extension of this interpolation grid
above Solar metallicity is the most uncertain part of our data analysis}.

\subsection{Results for the MDF}

Once the model grid is defined, we interpolate linearly within it,
in the manner described in Paper II, to estimate the metallicity
of each star.  Compared with
the previous analysis of Paper II, we have modified the interpolation code
to add minor improvements in the calculation of 
bolometric correction and the ability of the code
to use stars closer to the RGB tip.  For purposes of homogeneity,
we have therefore recalculated the MDFs for all three of our program
fields (8, 21, and 31 kpc) through the same updated routine.

As part of the interpolation routine we also compute an uncertainty
$\epsilon(Z)$ in the abundance for each star,
generated by the random photometric error
in magnitude and color at its position in the CMD.  
These uncertainties are determined numerically by simply adding
$\epsilon(I)$ and $\epsilon(V-I)$ to the given $(I,V-I)$ for the star
and recalculating $Z$.
In practice, the size of $\epsilon(Z)$ is determined almost
entirely by the scatter in color, $\epsilon(V-I)$, because the RGB 
evolutionary tracks are nearly vertical in the bolometric CMD and so small
shifts in magnitude alone do not affect the resulting MDF.

The data for our 8-kpc field, superimposed on the RGB model tracks, are
redisplayed in Figure \ref{cmd_bol} in the $(M_{bol}, (V-I)_0)$ plane within
which the interpolation process is actually carried out.  
The resulting histograms in [m/H], divided somewhat
arbitrarily into three luminosity bins, are shown in Figures 
\ref{mh2131_histo} and \ref{mh8_histo}.
Revisiting the combined MDF for the two outer fields (Figure \ref{mh2131_histo}),
we see in all bins a broad peak
near [m/H] $\simeq -0.5$, a steep decrease toward higher metallicity,
and a longer tail toward lower metallicity.  There are nearly
negligible numbers of stars either more metal-poor than [m/H] $\ltsim -1.5$
(which is, notably, the metallicity range including {\sl most}
of the halo stars in the Milky Way)
or more metal-rich than [m/H] $\gtsim -0.2$.  In the fainter luminosity bins,
the MDF becomes more spread-out
due to the increasing photometric scatter, but the peak location
and general distribution shape remain the same at all magnitudes.

For our 8-kpc field (Figure \ref{mh8_histo}) the patterns are not quite the same.
As we expected from the raw CMD, the populations in all the bins have 
higher proportions of metal-rich stars compared with their
outer-halo counterparts, but the fainter bins shift progressively
further to the metal-rich side.  This trend apparently violates the
requirement of internal self-consistency, i.e.~that the MDF for the
same population of stars should be independent of the luminosity range
sampled.  However, the highly consistent results from the outer fields 
(Figure \ref{mh2131_histo})
indicate that the problem is not an artifact of the interpolation routine
or the model grid itself.  A large part of the explanation is 
likely to be the detection incompleteness for extremely red stars,
which will affect the highest-luminosity bins somewhat more because the
RGB stars there (see Fig.~1) lie closer to the completeness cutoff line
than do equally metal-rich stars a magnitude further down the giant branch.

To take partial account of incompleteness effects, we recalculate
completeness-corrected MDFs for both the inner and outer fields
(where now each star is counted as $(f_Vf_I)^{-1}$, the
inverse of the photometric completeness at that point on the CMD; to
avoid wild excursions in the MDF we arbitrarily set the maximum correction
factor at 2.0).  From Figure \ref{mh8_histo},
it is evident that the metal-richer side of the distribution is enhanced
as a result, and more so in the higher-luminosity bins.
For the outer halo, incompleteness  is much less
important because the RGB does not reach far enough to the red to be
strongly affected.  
For $f-$values lower than $\simeq 0.5$, completeness effects become
poorly understood and very risky to apply.  
We regard the corrected histograms in Figure~\ref{mh8_histo}
only as an informed guess at the red-end shape of the MDF, and
not a substitute for raw data that reach appropriately deeper in the first place.

Lastly, we note that the fainter bins become progressively more
affected not only by photometric spread but also by unrecoverable
systematic errors:  higher and higher fractions
of the bluer stars are driven by 
photometric scatter out of the model grid entirely, but the stars 
driven to the red side by the same photometric random scatter stay within
the grid and skew the resulting MDF to higher metallicity. See Paper II
for more discussion.  In what follows, we restrict our discussion to
the brightest interval $M_{bol} < -2.5$ which is least affected by
this combination of errors and uncertainties.

Our final completeness-corrected histograms for the inner 
and outer fields are summarized in Table 1
and shown in Figure \ref{2histo}.  Regardless of the level
of completeness corrections, the inner-field MDF is clearly
more metal-rich on average, and extends to higher maximum metallicity,
than the outer halo.

\subsection{The MDF as a Probability Distribution}

As we discussed in Paper II, a physically natural way to display
the MDF which is easily connected to chemical evolution models, is
as the number of stars per unit heavy-element abundance, $dn/dZ$.
The numbers in Table 1 can readily be converted to this alternate
form, but while we are at it, we can also partially compensate for the
effects of random photometric errors.  Define 
$p(Z)$ as the probability of finding stars at a given abundance $Z$
as follows:

\begin{equation}
p(Z) \, = \, P_0 \, \sum_{i=1}^n \, {1 \over f_i} \cdot e^{-(Z-Z_i)^2/2\epsilon_i^2} \, . 
\end{equation}
In this expression, each star is replaced by 
a small Gaussian centered at its measured abundance
$Z_i$ and with dispersion $\epsilon_i$ equal to
the uncertainty in its calculated heavy-element
abundance $Z$. As before, $f$ is the detection incompleteness
at the star's location in the CMD.  The normalization constant $P_0$
is chosen to set $\int p(Z) dZ \equiv 1$ over all $Z$.

The probability distributions $p(Z) dZ$ for the inner and outer fields are
shown in Figure \ref{zdf}.  The curve for the outer fields is smooth and
featureless, reaching a 
peak at $Z$(max) $\simeq 0.25 Z_{\odot}$ (compare Fig.~13 from Paper II,
where basically the same material is plotted in raw histogram form).
By contrast, the curve for the inner field is extremely broad, declining
slowly from a peak at $\simeq 0.35 Z_{\odot}$ toward levels that extend
well above Solar abundance.  We emphasize
again, however, that its extension at high $Z$ is 
uncertain, as indicated by the difference between the completeness-corrected
and uncorrected curves shown in Figure \ref{zdf}.
We will use these distributions
as the basis for some brief modelling discussions in section 4 below.

\subsection{A First Look into the ``Bulge'' Population?}

In any representation, the MDF for our 8 kpc field  is plainly more
extended, more metal-rich, and more complex than the outer-halo MDF.
The inner MDF is likely to be broader 
because it is made up of a wider mixture of subcomponents from different
parts of the galaxy.
The effective radius of the bulge light is $r_e = 330''$
\citep{vdb76}, corresponding to $\simeq 6.4$ kpc. Our 8-kpc
field is threfore only $\sim 1.3 r_e$ from the center, and so 
a significant contribution from the inner bulge of the galaxy should
be present in our MDF.

If we now {\sl assume} that our inner-field MDF is made up of two
major components (bulge and halo), {\sl and} that the halo component
is similar to the outer-field MDF, we can subtract one from the other
to gain a first, albeit very rough, look at the MDF for the ``pure bulge''
component of NGC 5128.  To normalize the outer-field population to the
inner one, we use the metal-poor parts of the two MDFs where we can
plausibly expect the halo components to dominate. In Figure \ref{ratio_inout}, we plot
the ratio of numbers of stars in each field {\sl more metal-poor} than
a given cutoff metallicity [m/H].  For any cutoff in the range 
[m/H] $\ltsim -1$, this ratio stays roughly constant at 
$n_{in}/n_{out} \simeq 0.46 \pm 0.01$, indicating that the two MDFs
match up well over that range.  At higher metallicities, the
proportion of inner-field stars rises steadily and the two MDFs no longer
match one another.

In Figure \ref{diff_histo}, we show the result of 
subtracting one component from the
other.  The lower panel gives the residual, equal to 
($n_{in} - 0.46\cdot n_{out}$) (where both MDFs here are
the completeness-corrected versions).
The halo component ([m/H] $ < -1$) has now subtracted cleanly away, and
the remaining metal-rich part, by hypothesis,
represents the stars belonging to the outer bulge of the galaxy.
This residual MDF,
rising steadily from [m/H] $\simeq -0.8$ up to near-Solar abundance,
shows a broad peak and roughly symmetric shape, but as indicated above,
it probably still underestimates the numbers of stars with
above-Solar metallicity because of the inadequately understood
incompleteness effects of the photometry.

The formal mean and standard
deviation of the residual ``bulge'' MDF as shown here
are $\langle$m/H$\rangle = -0.19 \pm 0.01$ and $\sigma$(m/H) $= 0.25$ dex.
For comparison, \citet{tra00} find that the bulge metallicities
for a large sample of other elliptical galaxies, determined
from integrated-light spectral indices, lie in the typical range
[m/H] $\simeq -0.1$ to +0.3 with an average at +0.08
(we quote here their determinations for the integrated light within 
($r_e/2$); their ``core'' metallicities within ($r_e/8$) are slightly
more metal-enhanced at $\langle$m/H$\rangle = +0.26$).  Given 
the presence of a modest metallicity gradient, our
result for NGC 5128, measured further out in the bulge at $1.3 r_e$, 
appears to be quite consistent with their quoted range.  

A comparison with the Milky Way bulge is also possible.
The mean metallicity of the old Milky Way bulge stars within 
$r_{gc} \sim 1$ kpc (corresponding to about
half the classic effective radius) is at a very similar level:
for example, \citet{fro99} find 
$\langle$Fe/H$\rangle = -0.19 \pm 0.03$ for five fields
within $3^o$ latitude (0.4 kpc) along the minor axis. \citet{mcw94} find 
$\langle$Fe/H$\rangle = -0.27 \pm 0.12$ for 11 bulge giants at moderately
low latitude, and \citet{iba95} find that the MDF for four Milky Way
bulge fields peaks at [Fe/H] $\simeq -0.3$.  Thus our tentative findings for
the NGC 5128 `bulge'' fall very much in line with the ranges observed in
other elliptical and spiral bulges.

\section{Modelling the MDF}

The eventual aim of obtaining the metallicity distribution of the 
stellar population is to help place constraints on the
evolutionary history of NGC 5128.  Both theoretical modelling and
recent observations suggest that giant ellipticals are likely to
form or grow by a mixture of different processes including hierarchical
merging of gas clouds in the early universe 
\citep[e.g.][]{pea99,col00,dub98,kau93,the92,for97}; 
merging of previously formed large disk galaxies 
\citep[e.g.][]{bar92,hib99,naa01,sch87,ash92,zep00}; and nondissipative,
hierarchical merging of stellar fragments (e.g. C\^ot\'e et al.~1998, 2000, 2001).
Which of these processes dominates for any particular gE almost certainly
depends strongly on its environment and individual history.

Matching our data with some of these advanced modelling codes is far beyond
the scope of the present paper.  We will discuss
a specific comparison with one of the
semianalytical codes for hierarchical merging, GALFORM \citep{col00},
in a later paper \citep{bea02}.
For the present, in this section we briefly discuss 
a much simpler chemical evolution model 
capable of generating the first-order features of the MDFs that we observe.
The basic picture we adopt is one of hierarchical formation, in which
the bulk of the stars in the halo are assumed to form at relatively early epochs
within many gas-rich clouds that merged to build up the (eventual)
giant elliptical.  That is, the MDF is the result of a 
{\sl large number of individual star-forming events} within these smaller clouds
and within the growing potential well of the gE.  
This is almost the opposite view from what we adopted in the previous section,
i.e.\ that the MDFs that we see in these three fields 
resulted from a single series of {\sl in situ} star-forming
events, rather than from two rather distinct subcomponents.

\subsection{An Accreting-Box Model for the MDF}

As we argued in Paper II, perhaps the most interesting feature of the
MDF is the striking lack of low-metallicity stars in 
either the inner or outer halos.  A classic
closed-box model of chemical evolution with initial gas abundance $Z_0$ and 
nucleosynthetic effective yield $y$ 
leads to an MDF with a characteristic exponential-decay shape
$dn/dZ \sim e^{-(Z-Z_0)/y}$.  Without highly arbitrary adjustments
to the assumed star formation rates or the IMF, 
these basic models do not produce the steep rise in 
the number of stars at low $Z$ (see again Fig.~10,
and see Paper II and references cited there for
more discussion of the alternatives).

An alternate and almost equally simple
class of models is the ``accreting-box'' type, within which new gas is added
to the ``box'' after each round of star formation 
\citep[see, e.g.,][]{lar72,pag75,bin98,por98}.
A more intuitive way to state this is simply to visualize that
the many protogalactic clouds, which start with primordial unenriched
gas, are all busily forming stars within their own small potential wells
while they are simultaneously merging together.  That is, the merger 
epoch does not wait until after the protogalactic clouds have completed their
own local star formation, nor do the small clouds wait until they have
merged before beginning star formation.  To some degree,
both steps must be happening at once.

To quantify this approach in a straightforward way, 
assume \citep[following the notation
of][]{bin98, pag75} that the box has gas mass $M_g(t)$ and stellar mass
$M_s(t)$ at any time $t$,
and that star formation proceeds in discrete timesteps $\delta t$, within 
each of which a small fraction $e$
of the gas forms into stars. Of the mass $\delta M_s$ of newly formed stars, 
a fraction $\alpha$ stays locked up in stellar remnants and the remainder
$(1-\alpha)$ is returned to the interstellar medium, partially enriched
by the products of nucleosynthesis.  As usual, the 
yield $y$ is defined as the amount of 
heavy elements released back into the ISM per unit 
remnant mass.  By counting gains and losses in the mass $M_Z$ in
heavy elements, we eventually find \citep{pag75,bin98}
that the net change of the gas mass in heavy elements after each step
can be written as  
\begin{equation}
 \delta M_Z \, = \, -Z \cdot \delta M_s \, + \, (1-Z) y \cdot \delta M_s \, \simeq \, (y-Z) \cdot \delta M_s 
\end{equation}
where $\delta M_s = \alpha e M_g$ is the net gain in star mass.  In general, there
are no interesting analytic solutions to this relation (except
the artificial case $M_g = const$), but it can readily
be integrated numerically as soon as a prescription for the gas infall rate
is given.

To specify a form for the gas accretion rate that will be both realistic and
flexible enough to handle a variety of cases, we note that 
recent simulations of giant galaxy formation \citep[e.g.][]{the92,dub98,pea99,col00}
indicate that large numbers of gaseous fragments can be expected to 
come inward onto the central galaxy over time periods of
perhaps $\sim 2 $ Gyr give or take factors of two,
after which the infall rate gradually dies away.  
For our purposes, we parametrize this sequence as shown in
Figure \ref{infall_model}:  the rate of mass infall is assumed constant
for some initial period $\tau_1$, after which it falls off exponentially
with an e-folding time $\tau_2$:
$$ {\delta M \over  \delta t} \, = \, k \quad (t < \tau_1) \, ,$$ 
$$ \hskip 1.1 in = k \cdot e^{-(t-\tau_1)/\tau_2} \quad (t \ge \tau_1) \, . $$
An exponentially declining infall rate has also been used in the context
of disk formation by (e.g.) \citet{lac83,chi97}.
This ``delayed exponential'' model has four defining parameters 
in addition to the basic yield rate $y$:  these are the initial infall
rate $k$ and timescale $\tau_1$; the decay time $\tau_2$;
and the metallicity $Z_g$ of the infalling gas.  After 
time $t$, the total mass in stars will
be larger than the initial mass $M_0$ in the box by some factor 
$F \equiv M(t) / M_0$ which is determined by $k, \tau_1, \tau_2$.
The final mass $M_f$ of the model can then be set equal to the
observed mass of the present-day galaxy to estimate $M_0$, i.e.~the
initial ``seed'' mass of the proto-elliptical.

Although the step upward from the simpler closed-box model to an accreting-box
model inevitably involves a large increase in parameter space, it is
helpful that the different free parameters govern different parts
of the predicted MDF, as shown schematically in Figure \ref{sample_model}.  
The shape of the rising part of the MDF at low metallicity helps to fix
$k$ and $\tau_1$; the intermediate-metallicity section is determined
by $\tau_2$; and at late times and larger $Z$, the gas accretion has died
away to negligible levels and the system approaches closed-box evolution
fixed by the yield $y$.  The true closed-box model is a special case 
of our accreting-box model where $k=0$.

We assume, more or less arbitrarily, that the initial abundance of
the gas is primordial, $Z(0) = 0$.  We then adopt an abundance $Z_g$ for
the infalling gas and vary $k,
\tau_1$, and $\tau_2$ to obtain a match to the observed MDFs in Fig.~10.  To carry
out the numerical integration, we fix $M_0 = 1$ ``unit'' of initial gas
mass and take $e = 0.05$ (that is, 5\% of the available gas 
is converted to stars in each small timestep $\delta t$.  The exact choice
of $e$ is not critical; smaller fractions give a finer numerical mesh but
maintain the same final MDF shape).  After each timestep, the gas in the
box is assumed as usual to be promptly mixed, so that all 
the stars formed in the next timestep then have the same abundance $Z(t)$ 
as the surrounding gas. After the integration is complete, the amount of
stellar mass at each $Z$ is counted and the model MDF constructed.

\subsection{Model Results and Comments}

To match the steep observed rise of the MDF at low metallicity, we find
that the amount of gas accreted in each 
timestep initially needs to be comparable
with the amount converted to stars, i.e. $M_g({\rm added}) \sim e \cdot M_g$.
After a handful of such steps, however, the infall rate must begin dying away
to match the MDF peak and turnover region.  The parameters for representative
model solutions are summarized in Table 2, where $Z_g$ and $y$ are expressed
in Solar units, $\tau_1$ and $\tau_2$ are expressed in numbers of timesteps, and
the final galaxy mass $M_f$ is a multiple of the initial mass.
Figures \ref{enrich0} and \ref{enrich2} show these sample  results graphically.
In all cases the final galaxy mass $M_f$ 
is 2 to 4 times bigger than the initial $M_0$.  For the outer-halo fields,
the effective yield must be about one-quarter Solar, while for the inner halo,
it must be more than three times as high to generate the high$-Z$ section of the MDF.  

For any $Z_g \ltsim 0.2 Z_{\odot}$ it is possible to find models which
closely match the outer-halo MDF data.  However, for pre-enrichment values
much above $Z_g \gtsim 0.2 Z_{\odot}$, plausible fits become 
more difficult to find, because the incoming gas is already too enriched 
to produce the correct numbers of low$-Z$ stars.  We conclude, then, that
within the context of an accretion model and {\sl in situ} star formation,
the incoming gas needs to have quite low metallicity.

The inner-halo MDF is more easily matched with slightly more enriched 
gas ($Z_g \gtsim 0.1 - 0.2 Z_{\odot}$), but none of these simple models
can accurately reproduce the observed structure at $Z \gtsim 0.7 Z_{\odot}$.
If the excess of high$-Z$ stars is real, then
it may indicate that the metal-rich bulge component was created
separately such as by a late merger, where the accreted material was
already enriched.

The protogalactic clouds envisaged in our rough model correspond
strikingly with the characteristics of the
damped Ly$\alpha$ clouds. 
Recent high-resolution spectroscopic observations of these systems
over a large range of redshifts
\citep[e.g.,][]{pet00,pro99,pro00} have made it increasingly clear
that they are excellent candidates for protogalactic units 
out of which bigger galaxies assembled.  The observed heavy-element abundances
of these clouds cover a wide logarithmic range in [m/H] but 
are very metal-poor on average, with
$\langle$m/H$\rangle \simeq -1.6$ (cf.\ the references cited above),
corresponding to $\langle Z \rangle \sim 0.03 Z_{\odot}$.  This level
would place them in a regime bracketed by our Figures \ref{enrich0} and \ref{enrich2}.

The true mass infall rates cannot be derived directly from
our model, because the ``timestep'' $\delta t$ is arbitrary and must
be determined from external information. Unfortunately, the properties of the 
RGB stars which make 
them so useful as metallicity indicators also prevent them from 
supplying any very helpful information on their age distribution.  
Nevertheless, we can roughly calibrate the time sequence
if we forcibly {\sl assume} (following the 
hierarchical-merging simulations referred to above, as well as the 
evidence from the redshift distribution of the Ly$\alpha$ clouds) that the main
bulk of star formation took place over a plausibly short interval
such as $\sim 2$ Gy.
For all the models, we find that it takes typically $50 - 80$ 
timesteps with $e=0.05$ to reach 80\% of their final mass, 
suggesting $\langle \delta t\rangle \simeq 25-40$ Myr.

Finally, if the total stellar mass of NGC 5128 is
$M \simeq 4.1 \times 10^{11} M_{\odot}$ (for an integrated magnitude
$V^T = 6.2$ \citep{vdb76}, a distance modulus $(m-M)_V = 28.3$, and
$(M/L)_V = 7$), the amount of gas converted to stars
in each timestep can be translated to a star formation rate (SFR)
knowing $\delta_t$.  The corresponding average SFR in the
first 2 Gy and integrated over the entire galaxy
is $\simeq$ 150 $M_{\odot}$ y$^{-1}$.  The 
{\sl maximum} SFR for each model is listed at the end of 
Table 2.\footnote{Strictly speaking, for the outer-halo fields we should
define $M_f$ as the total mass of only the halo and not the entire 
galaxy, and similarly for the outer-bulge field we should use only the
total mass of the bulge.  However, the end result for the maximum SFR
added up over the entire galaxy is the same in either case.}
SFRs this large are an order of magnitude
higher than observed in large galaxies at redshifts $z \sim 1$
\citep[e.g.,][]{pos01}, but are smaller than the rates observed
in various extreme starburst systems such as the Ultra-Luminous Infrared Galaxies
\citep{ana00,dey99,sco01,alo98}.

\section{Globular Clusters and a New ``Specific Frequency Problem''}

As a final part of our discussion, we return to 
an issue first raised in Paper I:  how does
the metallicity distribution of the field-halo stars compare with
that of the globular clusters in NGC 5128?  Eventually, we would like
to encompass the MDFs of both field-halo stars and clusters
within a single formation history.  To address this question,
we take the samples of individually determined 
cluster metallicities published by \citet{har92} (from the Washington
$(C-T_1)$ index) and by \citet{rej01} (from the $(U-V)$ color index),
each of which contain 60 to 70 clusters drawn from a wide range of radial distances
in the halo.  We view this as only a preliminary comparison,
since considerably larger samples of clusters will soon be 
available from other wide-field imaging projects now in progress.
Nevertheless, even with this rough comparison an interesting trend emerges.

\subsection{Calibration of $(C-T_1)$ Metallicity Index}

The Washington color $(C-T_1)$ is an effective metallicity index
for globular clusters \citep{gei90} and
has been used as such in many studies.  However, the original
calibration of $(C-T_1)$ in terms of cluster metallicity employed
Milky Way cluster reddenings and [Fe/H] values which 
have been extensively revised and updated over the intervening years,
to the point where it is of interest to redo the calibration.
For the Milky Way cluster $(C-T_1)$ colors we use the fundamental
list of \citet{har77}, while for their reddenings and modern [Fe/H]
values we use the 1999 edition of the \citet{har96} catalog, along
with the reddening conversion $E(C-T_1) = 1.966 E(B-V)$ \citep{har79}.
Combining this material for 48 clusters yields the relation shown
in Figure \ref{washington}.  

With the contemporary data, the correlation between the two now
appears to be mildly nonlinear.  A quadratic least squares solution
of $(C-T_1)_0$ against [Fe/H]
(ignoring two obviously discrepant points at intermediate color, which
belong to NGC 288 and 6522) yields
\begin{equation}
(C-T_1)_0 \, = \, 1.998\, + \, 0.748 {\rm [Fe/H]} \, + \, 0.138 {\rm [Fe/H]}^2
\end{equation}
while the inverse solution gives
\begin{equation}
{\rm [Fe/H]} \, = \, -6.037 \, \big (1 - 0.82 (C-T_1)_0 + 0.162 (C-T_1)_0^2\big ) \, .
\end{equation}
These reproduce the data with an rms scatter of $\pm 0.16$ dex in [Fe/H],
or $\pm 0.055$ mag in $(C-T_1)_0$.  As Figure \ref{washington} shows, these two 
curves are almost identical except at the extreme low-metallicity end,
which is not well constrained by the data and where in any case the
color index becomes insensitive to metallicity.
We use this new calibration to convert the $(C-T_1)$ indices of the
NGC 5128 clusters \citep{har92} into metallicity.  

For the \citet{rej01} sample, we use the correlation between $(U-V)_0$
and [Fe/H] derived by \citet{ree94},
\begin{equation}
{\rm [Fe/H]} \, = \, -3.061 \, + \, 2.015 (U-V)_0 \, .
\end{equation}

The metallicity indices for the {\sl globular clusters} are calibrated
in terms of [Fe/H], while our data for the {\sl field stars} are in terms
of [m/H] = log $(Z/Z_{\odot}) \simeq$ [Fe/H] + [$\alpha$/Fe].  Recent
compilations of spectroscopic abundance measurements for old-halo giants
in many galaxies \citep[e.g.][]{she01} indicate that the 
[$\alpha$/Fe] ratio has significant star-to-star scatter and that its mean value
increases slowly with decreasing metallicity, staying
in the range $0.2 - 0.3$ for $-3 \ltsim$ [Fe/H] $\ltsim -0.5$.  At more
nearly Solar metallicity, spectra of Milky Way bulge stars and of  
the integrated bulge light of other large galaxies \citep{mcw94,tra00}
indicate that [$\alpha$/Fe] remains near $+0.2$ though again with
considerable object-to-object variance.  Although it is not entirely
clear whether this rather mixed ensemble of evidence should apply to
NGC 5128 in particular, we adopt it as the best present guess.
In summary, to convert our stellar data from 
[m/H] to [Fe/H] we adopt
[$\alpha$/Fe] = +0.2 for [Fe/H] $\geq -0.6$, and then use a linear 
increase of [$\alpha$/Fe] from +0.2 to +0.3 as [Fe/H] decreases from
$-0.6$ to $-2.0$.

\subsection{Comparing Clusters With Field Stars}

The available NGC 5128 globular cluster data include objects with projected
galactocentric distances from $R \simeq 2'$ to almost $22'$, and 
although the combined sample comprises only 127 clusters, it is already
clear that the outer-halo objects are more metal-poor on average.
More or less arbitrarily, we have divided the total set of clusters
into two roughly equal subsets by radius (within $R=8'$ and beyond $R=8'$)
for comparison with the inner-halo and outer-halo field star samples,
leaving about 65 clusters in each group.

The results are shown in Figure \ref{feh2}.  The histograms for the clusters have
been calculated with a smoothing kernel of 0.1 dex, while the stellar
histograms have been smoothed by 0.05 dex, consistent with the internal
uncertainties resulting from the random errors of photometry.  
In each graph, both curves have been normalized to the same total 
population of objects.  The distinct lumpiness of the two cluster MDFs
is a plain result of the still-small sample size and should not be
ascribed any particular importance.
For the inner halo, we find that the clusters and
field stars have basically similar MDFs, with an extremely wide range of metallicities
and broad peaks near [Fe/H] $\sim -0.5$.  Given that the two distributions were
established in somewhat different ways ($(V-I)$ colors of RGB stars, versus 
$(C-T_1)$ integrated colors of clusters), we suggest that there is no strong
reason to claim that they are fundamentally different.

For the outer halo, however, the clusters and field stars are
strikingly different (in a formal statistical sense, they
are different at the $> 99.9$\% confidence level from a K-S test).
The cluster MDF at least roughly resembles what we find
for the globular clusters in the Milky Way and M31 
\citep[e.g.][]{bar00,har01}, with a primary metal-poor peak at
[Fe/H] $\simeq -1.6$ and a secondary group of more metal-rich
clusters extending up to roughly Solar abundance.  Unlike the
Milky Way, however, the outer-halo stars lie at a considerably
higher mean metallicity level, corresponding to about one order of magnitude in   
heavy-element abundance $Z$.  The case for resemblance to M31
is much stronger:  as pointed out by \citet{dur01} and \citet{hh01},
the same dichotomy between the clusters and field-halo stars appears
there, and to very much the same degree.

\subsection{Specific Frequency:  Discussion and Speculation}

The ratio of numbers of globular clusters to field stars is defined
as the specific frequency $S_N$ \citep{hvdb81,har01}.  {\sl If} it is reasonable
to assume that the globular clusters at a given [Fe/H] can be associated
with the field stars at the {\sl same} metallicity, then
our NGC 5128 data offer the chance to plot $S_N$ 
in a new way, as a function of the {\sl metallicity} of the subpopulation.  
We did this roughly in Paper I by breaking the stars and clusters into
just two metallicity groups, but the more comprehensive
datasets now allow us to divide the metallicity range a bit further.
Results for both the inner and outer halo, and
for five metallicity bins in each zone of width $\Delta$[Fe/H] = 0.4 dex,
are listed in Table 3 and plotted in Figure \ref{sn_feh}.  In each bin, we integrate
under the smoothed MDFs in Fig.~18 for the given metallicity range to
estimate the relative numbers of stars and clusters.
The resulting ratios 
are normalized so that the total $N_{cl}/N_{\star}$  over all metallicities equals
the specific frequency $S_N$(total) = 2.6 for the whole galaxy (see Paper I).
Evidence from wide-field starcounts \citep{har84} suggests that the
globular cluster system follows the same overall radial
distribution as the integrated halo light and thus the same global
specific frequency can be used for both regions.

The actual zeropoints for each region are, however, less important than
the trend with metallicity:  we find that $S_N$ is a strong function of [Fe/H],
changing by roughly a factor of three over the entire metallicity range.
(The anomalously high $S_N$ for the most metal-rich
clusters in the outer halo is likely to be strongly
affected by small-sample statistics and is thus not very significant.)
For [Fe/H] $\gtsim -1.2$, we find $S_N$ values near $1-3$, which are typical
of those in field ellipticals, spirals, and recent merger products of disk
galaxies \citep{har01}.  For [Fe/H] $\ltsim -1.2$, the ratio is $4-8$,
comparable with many giant E galaxies, or
(see below) many of the smallest dwarf ellipticals.

Metallicity-based differences in $S_N$ have been hinted at in previous
work for other large galaxies such as M31 and NGC 4472
\citep[see][as well as Paper I for NGC 5128 itself]{dur94,for97,har01}, 
but the existence of more extensive
MDFs for both clusters and halo stars in the same system now
allows us to quantify the trend much further.  It suggests either
that low-metallicity clusters formed at considerably higher efficiency
than metal-richer ones, or that in some other way
the formation of globular clusters did not go in lockstep with the field
stars.  This disparity is all the more puzzling, given
the strong evidence for a global cluster formation efficiency of
about 0.25\% by mass that is remarkably similar 
in large galaxies of all types and sizes \citep{mcl99,bla99,kav99,har01}.

For globular cluster systems there are at least three different
kinds of ``specific frequency problems''.  The first
of these, uncovered more than two decades ago, is the correlation of 
$S_N$ with E galaxy luminosity and the existence of central-supergiant
galaxies (M87 and others like it) with strikingly large cluster populations.
A plausible solution to this issue \citep{mcl99,bla99,kav99} 
is that the $S_N$ values become much more nearly uniform when normalized to
the total masses of the host galaxies including their 
X-ray halo gas.  A second and almost equally old ``specific frequency problem''
is the overall large range in $S_N$ amongst rich cluster ellipticals,
field ellipticals, and spirals \citep[see][for a review]{har01}.
Understanding the origin of this range and variety
is likely to involve numerous details of their formation 
histories including environmental factors such as the tendency of
spiral/spiral mergers to form low$-S_N$ ``field'' ellipticals.  

The correlation between $S_N$ and metallicity which we raise
here is a third fundamental
kind of ``specific frequency problem'', and presents a new challenge to our 
understanding of the link between globular clusters and halo stars.
Much observational and theoretical evidence indicates that (a) star clusters
form within giant molecular clouds (GMCs), with perhaps a handful of clusters
forming within any one GMC, and that (b) the typical
masses of the protoclusters are about 3 orders of magnitude smaller
than the total mass of the parent GMC itself \citep{hp94,mp96,mcl99}.  
The mean mass ratio $\epsilon_{cl} = 0.0025$ derived by \citet{mcl99}
corresponds to a baseline specific frequency $S_N \sim 5$ and
is, in this hypothesis, the average mass ratio $M_{cl}/M_{GMC}$ of clusters
formed within a typical GMC.  Star clusters with 
masses $10^4 - 10^6 M_{\odot}$
(that is, globular clusters) should then form within GMCs of masses
$10^7-10^9 M_{\odot}$.  GMC masses in this range make them
easily identifiable with protogalactic subsystems such as
the Ly$\alpha$ clouds \citep{bur01}, the supergiant molecular clouds
of \citet{hp94}, or the small cool-gas disks of \cite{col00} which merge
hierarchically to build larger galaxies at the same time as they are
forming their own embedded star clusters.

As did \citet{for97}, \citet{har01}, and \citet{bur01}, we imagine 
a sequence of events in which the first major round of globular cluster 
and star formation took place within these
dwarf-galaxy-sized gas clouds, scattered
throughout the much larger potential well of the giant
proto-elliptical.  These first clusters and field stars
took on the low metallicities
of their primordial host clouds.  As the clouds merged hierarchically, 
later continuous
rounds of star formation took place in which the progressively
more enriched gas formed the metal-richer clusters and field stars.
But some additional feature of the story
is needed to force the ratio of field stars to globular clusters steadily
upward (or conversely, to push $S_N$ steadily downward) in these later
rounds of star formation, as the evidence of Fig.~19 would indicate.

We suggest that this additional feature may be
the relative timing of globular cluster formation {\it versus} their associated
field stars.  If the clusters, which should form out of the densest, most massive,
or highest-pressure clumps of gas within their host GMCs, 
are constructed {\sl soonest} in the sequence -- that is, ahead of the majority
of the field stars -- then the
effective $S_N$ emerging from the entire GMC will depend on when the
star formation history is truncated.  Earlier versions of this idea were
suggested by \citet{dur96} and \citet{har01}.

The essential concept is illustrated schematically in Figure \ref{schematic}.
Within a given protogalactic cloud (supergiant GMC), the clusters are
envisaged to form predominantly near the beginning of the GMC's
history.  By contrast, the general star formation 
distributed everywhere throughout the cloud in less 
dense local regions and at lower conversion efficiency 
is assumed to take longer to ramp up and then die away.
A ``normal'' specific frequency (number of clusters per unit field-star mass)
would result if the host GMC is able to complete its full sequence of
star formation (in Fig.~20, up to time T1).  However, {\sl if the normal
sequence is truncated at an earlier time} $T2$ -- for
example, if the GMC itself is disrupted and its remaining gas ejected -- then
the net specific frequency can be considerably larger:  the same number
of globular clusters is still present, having formed very early on,
but most of the general stellar population does not get a chance to form.

This premature truncation of the internal enrichment and star formation
is exactly what would be expected for isolated
protogalactic clouds of masses $\ltsim 10^9 M_{\odot}$:  as first calculated
by \citet{dek86}, the initial major round of supernovae and stellar winds
in such dwarf-galaxy-sized systems 
would be sufficient to drive the rest of the gas out of their modest
potential wells and terminate any further star formation.  A prediction
emerging from this combined picture is that dwarf
ellipticals, which can plausibly be viewed to have evolved from single
protogalactic clouds in isolation, should have progressively larger specific
frequencies for
smaller dwarfs (in addition, of course, to the long-established 
observational trend that smaller dwarfs should have lower mean metallicity).
The available $S_N$ data for dE galaxies \citep{dur96,mil98} are strongly
consistent with this prediction, as shown by \citet{mcl99}. The smallest dE,N systems have
$S_N$ reaching the remarkably high range $\sim 15 - 25$ (though with considerable
scatter), while the largest dwarfs have $S_N$ more consistently at ``normal'' levels of
$\sim 3-6$ \citep[see the discussions of][]{mcl99,har01}.
Another piece of observational evidence that is, perhaps, related to this
picture is that very high specific frequencies or specific luminosities
have been derived for some starburst galaxies 
\citep[e.g.,][]{meu95,zep99,lar00}.  The cluster-to-field-star ratio 
very early in the burst will look nominally larger than normal if the
clusters form preferentially earlier.  Nevertheless, it is intriguing that
most such systems \citep[see][for a compilation]{lar00} have specific
luminosities at the 1\% level or less, comparable with the ``universal''
0.25\% efficiency ratio for old cluster systems.

We suggest that the high $S_N$ levels exhibited at {\sl low metallicities
in giant galaxies} 
(Fig.~19 and the discussion above) may be  the result of the
same basic phenomenon.  At early times, the protogalactic dwarfs were
dispersed widely enough that they could, for a short time,
evolve almost in isolation.  They formed
metal-poor clusters and some stars, but then ejected most of their gas, leaving
behind a low-metallicity but high$-S_N$ stellar population.  The large amount
of leftover gas then re-collected in the much bigger potential well of
the giant galaxy as hierarchical merging continued, so that in the later rounds
of star formation most of the gas could be kept {\sl in situ} and the star formation
could run to completion, yielding a normal ratio of clusters to field stars.

In this admittedly speculative picture, the overall level of heavy-element
enrichment is still closely connected with chronology 
within any one GMC.  If the later rounds of star formation behave the
same way as the first round (that is, if globular clusters form soonest
within any GMC regardless of its initial metallicity),
then we would expect that {\sl the highest-metallicity
field stars in the bulge of the galaxy should be more metal-rich than
the highest-metallicity globular clusters}.  In other words, the MDF for
the field-star population should extend to a higher ``top end'' than
that of the clusters.  In NGC 5128, it should be possible to test this
prediction with a more comprehensive set of cluster data, along with
efforts to extend the field-star MDF to higher metallicity limits than
we were able to do in the present study.

If the timing argument we have outlined above proves to be wrong,
then the obvious alternative is to invoke an intrinsic
dependence of cluster formation efficiency on metallicity. That is, 
we would in that case assume that the very low$-Z$ protogalactic clouds 
inhabiting the early universe converted their gas into dense, bound
star clusters about 3 times more efficiently than did the later
generations of GMCs which had been enriched to levels higher than 
about one-tenth Solar abundance.  A third possibility, recently revived
by \citet{cen01}, would be to assume arbitrarily that the
lowest-metallicity clusters were formed in the early {\sl pregalactic} universe
by way of a fundamentally different route 
(in Cen's model, at the epoch of cosmological reionization).  This latter
route appears to us to be much less attractive since it does not relate
to any currently visible evidence for star and cluster formation in GMCs,
nor does it explain the otherwise-similar properties of globular 
clusters (mass distribution functions, structural parameters) 
at all metallicities and in all types of galaxies.

\section{Summary}

We present new HST/WFPC2 photometry in $(V,I)$ for a starfield 8 kpc from the
center of the nearby giant E galaxy NGC 5128.  By using a grid of red-giant
evolutionary tracks finely spaced in metallicity, we convert the color-magnitude
diagram of the stellar population into a metallicity distribution function,
and combine it with the MDFs for our previously analyzed outer-halo 
fields at 21 and 31 kpc distance (Papers I and II).

In all three fields, the CMD is completely dominated by a conventionally
``old'' (many Gy) stellar population of red giants, with moderately
high mean metallicity $\langle$m/H$\rangle \sim -0.5$ and very few
low-metallicity stars like the ones dominating the Milky Way halo.
However, the innermost of our three fields is distinctly more metal-rich
than the outer ones; if we arbitrarily view it as consisting of two
sub-populations (a halo component like the outer fields, plus a bulge component), 
then the ``bulge'' by itself has a mean metallicity [m/H] $\simeq -0.2$,
similar to bulge compositions in many other large elliptical and
spiral galaxies.

We find that a simple accreting-box chemical evolution model that assumes 
a gas infall rate that is initially rapid but then declines
exponentially with time is capable
of closely matching the MDF of the outer-halo fields, as long as
the infall gas has low metallicity ([m/H] $\sim -1$ or less) and
the effective nucleosynthetic yield is $y \sim 0.3 Z_{\odot}$.
The same type of model with $y \sim 0.8 Z_{\odot}$
can also match the inner-halo field.

Finally, we compare our field-star MDFs to the MDF for
the old globular clusters in NGC 5128, with both samples divided into inner
and outer regions.  In both regions, there is a striking increase of
net specific frequency (number of clusters per unit halo field-star numbers)
at low [Fe/H], with the lowest-metallicity bins having 3 times more clusters
per unit halo light than the highest-metallicity bins.  We suggest that
a way to explain this trend, within the context of a hierarchical-merging
formation picture, is that globular clusters form preferentially sooner
within any one starburst than the accompanying field stars.  The
first clusters to form were in relatively isolated, small, low$-Z$ protogalactic
clouds, which ejected most of their gas after the first round of star
formation, leaving behind a population of low-metallicity clusters and
only a few field stars.  In the later rounds of star formation after
the merging process had advanced much further, the sites of star and cluster
formation were in the much deeper potential well of the emerging giant
elliptical.  They could thus hold onto much higher fractions of their gas, allowing 
star formation to run to completion and producing proportionally larger numbers
of high-metallicity stars for each globular cluster.

\acknowledgments

This work was supported by the Natural Sciences and Engineering
Research Council of Canada through research grants to the authors.
We are pleased to acknowledge the hospitality and support at 
Mount Stromlo Observatory (RSAA/ANU)
during the authors' research leaves when this paper was written.


\clearpage


\begin{deluxetable}{rrrcrrr}
\tablecaption{Completeness-Corrected Metallicity Distribution Functions\label{tab1}}
\tablewidth{0pt}
\tablehead{
\colhead{[m/H]} & \colhead{n(Outer)} & \colhead{n(Inner)} &\colhead{$\quad$}
& \colhead{[m/H]} & \colhead{n(Outer)} & \colhead{n(Inner)} \\
}
\startdata
 -2.65 &     2.1 &     2.1 &&  -1.05 &   210.7 &   113.1 \\
 -2.55 &     7.3 &     2.1 &&  -0.95 &   282.4 &   135.7 \\
 -2.45 &     3.2 &     5.2 &&  -0.85 &   309.1 &   175.4 \\
 -2.35 &     5.2 &     2.1 &&  -0.75 &   407.9 &   235.0 \\
 -2.25 &    15.7 &     3.1 &&  -0.65 &   530.9 &   320.3 \\
 -2.15 &    14.6 &     6.2 &&  -0.55 &   549.5 &   428.7 \\
 -2.05 &    18.8 &    11.5 &&  -0.45 &   614.2 &   497.7 \\
 -1.95 &    27.1 &    10.4 &&  -0.35 &   467.7 &   495.4 \\
 -1.85 &    37.6 &    15.7 &&  -0.25 &   332.1 &   464.6 \\
 -1.75 &    40.7 &    16.7 &&  -0.15 &   148.3 &   429.5 \\
 -1.65 &    42.0 &    23.1 &&  -0.05 &    25.4 &   304.2 \\
 -1.55 &    49.2 &    23.0 &&   0.05 &     1.9 &   340.6 \\
 -1.45 &    74.2 &    38.7 &&   0.15 &     0.0 &   160.6 \\
 -1.35 &    87.8 &    23.0 &&   0.25 &     0.0 &    72.0 \\
 -1.25 &   111.7 &    56.5 &&   0.35 &     0.0 &    40.0 \\
 -1.15 &   170.7 &    81.7 &&   0.45 &     0.0 &     4.0 \\
 \enddata

\end{deluxetable}

\clearpage

\begin{deluxetable}{rcccc}
\tablecaption{Fitting Parameters for Accreting Box Models\label{tab2}}
\tablewidth{0pt}
\tablehead{
\colhead{} & \colhead{Outer} & \colhead{Inner}
& \colhead{Outer} & \colhead{Inner} \\
\colhead{Parameter} & \colhead{Fields} & \colhead{Field}
& \colhead{Fields} & \colhead{Field} \\
}
\startdata
$Z_g/Z_{\odot}$  & 0.0  &  0.0  &  0.2  &  0.2  \\
$y/Z_{\odot}$    & 0.32  & 0.87  & 0.25  &  0.81 \\
$\tau_1/\delta_t$   &  7  &  5  &  1  &  0   \\
$\tau_2/\delta_t$   & 35  & 20  & 19  &  12  \\
$M_f/M_0$  & 3.5  & 3.9  & 1.9  &  2.4  \\
max SFR ($M_{\odot}/y$) & 225 & 155 & 240 & 158 \\
 \enddata

\end{deluxetable}

\begin{deluxetable}{rcc}
\tablecaption{Specific Frequency vs. Metallicity \label{tab3}}
\tablewidth{0pt}
\tablehead{
\colhead{[Fe/H] Range} & \colhead{$S_N$ (inner halo)} & 
\colhead{$S_N$ (outer halo)} \\
}
\startdata
$ < -1.6$ & $3.4 \pm 1.2$ & $8.3 \pm 1.9$ \\
$(-1.6, -1.2)$ & $4.4 \pm 1.5$ & $4.9 \pm 1.2$ \\
$(-1.2, -0.8)$ & $4.0 \pm 1.1$ & $1.3 \pm 0.4$\\
$(-0.8, -0.4)$ & $1.6 \pm 0.4$ & $0.9 \pm 0.3$ \\
$ > -0.4$ &  $ 1.6 \pm 0.4$ & $10.3 \pm 3.1$ \\
 \enddata

\end{deluxetable}

\clearpage


\begin{figure}
\caption{Color-magnitude diagram for 17,326 measured
stars in the PC1 field, centered a projected distance of 8 kpc southwest
of the center of NGC 5128.  The dashed line indicates the magnitude limit
of 50\% detection completeness (see text for explanation).
\label{fig1}}
\end{figure}

\begin{figure}
\caption{Artificial-star tests for the DAOPHOT/ALLSTAR photometry
in the PC1 field.  Stars are added 500 at a time to the original image
with $(I,V-I)$ values along the dispersionless line (left panel).
Their measured values (right panel) scatter symmetrically about the
input relation.  The dashed line shows the 50\% detection completeness
boundary.  
\label{fig2}}
\end{figure}

\begin{figure}
\caption{The two panels show, for the artificial-star tests, the
scatter of measured magnitudes versus input magnitude ($\Delta I$ versus $I$,
$\Delta V$ versus $V$).  Although the random uncertainty increases with
magnitude, no systematic bias is present over the upper magnitude range used
for the definition of the metallicity distribution function (see text).
\label{fig3}}
\end{figure}

\begin{figure}
\caption{Detection completeness of the photometry as a function of
magnitude, derived from the artificial-star tests.  Here the completeness
fraction $f$ is defined as the number of measured stars in a given 
magnitude bin relative to the number of input stars in that bin.  The
two smooth curves are Pritchet interpolation functions \citep{fle95}.
The curves shown are for mean colors $(V-I) = 2$ and are only to be taken
as indicative of the average magnitudes at which the photometry becomes
severely incomplete. 
\label{fig4}}
\end{figure}

\begin{figure}
\caption{Calibration grid for the derivation of stellar metallicities.
The open circles indicate fiducial points for the seven ``standard''
Milky Way clusters described in the text (the three most metal-rich,
NGC 104, 6553, and 6791 are labelled).  The solid lines are the evolutionary
tracks for the $\alpha-$enhanced models of VandenBerg et al.\ (2000) used
in Paper II, while the two dashed lines are estimated tracks for heavy-element
abundances $Z = 0.025$ and 0.045 as described in the text.
\label{calibration}}
\end{figure}

\clearpage
\begin{figure}
\caption{Color-magnitude array for the 8 kpc field, plotted in terms
of bolometric magnitude versus intrinsic color.  The RGB model grid
from VandenBerg et al.~(2000; solid lines) 
are superimposed on the data.  The two dashed lines are two more metal-rich
tracks defined as described in the text.  Interpolation within the grid is used
to estimate the metallicity of each star.
\label{cmd_bol}}
\end{figure}

\begin{figure}
\caption{Metallicity distribution function (MDF) for the combined outer-halo
fields (21 kpc and 31 kpc) discussed in Papers I and II.  The results have
been recalculated from the stellar model grid of Figure 5
through the updated interpolation code as described in the
text.  The MDFs for four different luminosity intervals along the RGB are
shown, from the top of the RGB down to a point two magnitudes lower.
Note the increased spread at fainter levels, due to increased photometric
scatter.  The unshaded histograms show the completeness-corrected results
as described in the text.
\label{mh2131_histo}}
\end{figure}

\begin{figure}
\caption{MDF for the inner-halo 8 kpc field, calculated in the same way
as for the outer-halo fields in the previous figure.  The shaded histograms
represent the raw number counts, while the higher unshaded histogram line
represents the totals corrected for detection incompleteness as described
in the text.
\label{mh8_histo}}
\end{figure}

\begin{figure}
\caption{Final MDFs for the outer-halo fields (upper panel) and the
inner-halo field (lower panel).  Table 1 contains the same data, tabulated
in 0.1 dex intervals.  All RGB stars brighter than $M_{bol} = -2.5$,
i.e. the upper 1.5 magnitudes of the RGB, are used here in an unweighted sum.  
In each panel the shaded histogram is corrected for photometric incompleteness,
while the solid line underneath shows the raw, uncorrected total.
\label{2histo}}
\end{figure}

\begin{figure}
\caption{Heavy-element abundance distribution for the inner-halo (solid line)
and outer-halo (dashed line) fields, presented as a probability distribution
as defined in the text.  Here, $p(Z) dZ$ is the probability 
of finding a star with metallicity in the range $(Z, Z+dZ)$ and the 
integral over all $Z$ is normalized to 1.000.  In this particular
scaling, the interval $dZ$ is equal to $0.01 Z_{\odot}$.  The {\sl dotted line}
at bottom shows the distribution for the inner field without any correction
for photometric incompleteness.
\label{zdf}}
\end{figure}

\clearpage
\begin{figure}
\caption{Normalization of the metal-poor section of the MDF in
the inner and outer fields.  Here n({\it inner})/n({\it outer}) is
the ratio of the numbers of stars in each field with [m/H] {\sl less}
than the given cutoff value.  For [m/H]  $\ltsim -1$ the ratio is nearly
constant (see text).
\label{ratio_inout}}
\end{figure}

\begin{figure}
\caption{{\sl Upper panel:} The MDFs for the inner (unshaded histogram)
and outer (shaded histogram) regions of the NGC 5128 halo.  The outer-halo
MDF has been normalized to match the inner-halo one for the metal-poor
range [m/H] $< -1$ defined in Figure 11.  {\sl Lower panel:} The residual
MDF after subtraction of the scaled outer-halo region from the inner-halo one.
This differential MDF by hypothesis represents the outer bulge stellar
population of NGC 5128.  The roughly normal distribution of the residual MDF
is indicated by the Gaussian curve with mean at [m/H] $=-0.19$ and
standard deviation 0.25 dex.
\label{diff_histo}}
\end{figure}

\begin{figure}
\caption{The ``delayed exponential'' gas infall for the accreting-box
model used in the text.  Gas is assumed to fall in at a constant
rate $k$ for a time $\tau_1$, then its infall rate declines exponentially
with an e-folding time $\tau_2$.
\label{infall_model}}
\end{figure}

\begin{figure}
\caption{A schematic MDF resulting from an accreting-box model (solid line).
The parameters $k$ and 
$\tau_1$ determine the height and extent of the initial rise at lowest
metallicity; $\tau_2$ determines the shape of the turnover region at
the peak; and the yield rate $y$ determines the steepness of the decline
at high metallicity.  The closed-box model (dashed line) corresponds
to $k=0$, and follows an exponential decline from the start.
\label{sample_model}}
\end{figure}

\begin{figure}
\caption{Sample fits of the accreting-box chemical evolution
model to the inner- and outer-halo MDFs.  We use here the 
probability distribution form of the MDF in terms of linear
abundance $Z$.
{\sl Solid lines} show the observed MDFs from Figure 10, while
the {\sl dashed lines} show the models with parameters as
listed in Table 2.  These models assume the accreted gas has
``primordial'' abundance, $Z_g = 0$.
\label{enrich0}}
\end{figure}

\clearpage
\begin{figure}
\caption{Same as the previous figure, but for accreted gas with abundance
$Z_g = 0.2 Z_{\odot}$ ([m/H] $= -0.7$).
\label{enrich2}}
\end{figure}

\begin{figure}
\caption{Calibration of the $(C-T_1)_0$ integrated color against
metallicity, for 48 Milky Way clusters.  The solid line is the
quadratic relation from Equation 5 given in the text, the dashed line
is from Equation 6, and the short-dotted line
is the original linear relation of Geisler \& Forte (1990).
\label{washington}}
\end{figure}

\begin{figure}
\caption{Comparison of the metallicity distributions in [Fe/H] for
the field halo stars (dashed lines) and NGC 5128 globular clusters
(solid lines).  Each distribution has been smoothed as described in
the text.  The {\sl upper panel} shows the 8-kpc halo field compared
with the globular clusters at projected distances less than 8 arcmin
from the center of the galaxy.  The {\sl lower panel} shows the
two outer-halo fields (21 kpc and 31 kpc) compared with the globular
clusters outward of 8 arcmin.
\label{feh2}}
\end{figure}

\begin{figure}
\caption{Specific frequency $S_N$ of the globular cluster population
in NGC 5128, plotted as a function of metallicity interval [Fe/H].
The data points are listed in Table 3, separately for the inner-halo
($R < 8'$) and outer-halo ($R > 8'$) regions.  The dashed line at
$S_N = 2.6$ is the global average for the entire galaxy.
\label{sn_feh}}
\end{figure}

\begin{figure}
\caption{Schematic concept for star and cluster formation within a
single GMC or pregalactic gas cloud.  
(In reality the formation rates would be
much more segmented and ``burst''-driven than the smooth curves shown
here.)  If the gas supply in the cloud continues star
formation till time T1 (dashed line at right)
a normal ratio $S_N$ of clusters to field stars will result.  However, if the
formation is truncated at some earlier time T2, then the cloud will
end up with a high effective specific frequency.
\label{schematic}}
\end{figure}

\end{document}